\documentclass[a4paper,11pt]{amsart}
\usepackage{graphicx}
\begin{document}
\hyphenation{gra-vi-ta-tio-nal re-la-ti-vi-ty Gaus-sian
re-fe-ren-ce re-la-ti-ve gra-vi-ta-tion Schwarz-schild
ac-cor-dingly gra-vi-ta-tio-nal-ly re-la-ti-vi-stic pro-du-cing
de-ri-va-ti-ve ge-ne-ral ex-pli-citly des-cri-bed ma-the-ma-ti-cal
de-si-gnan-do-si coe-ren-za pro-blem gra-vi-ta-ting geo-de-sic
per-ga-mon cos-mo-lo-gi-cal gra-vity cor-res-pon-ding
de-fi-ni-tion phy-si-ka-li-schen ma-the-ma-ti-sches ge-ra-de
Sze-keres opi-nion}
\title[Einsteinian Manifolds and Gravitational Waves]
{{\bf Einsteinian Manifolds\\and Gravitational Waves}}

\author[Angelo Loinger]{Angelo Loinger}
\address{Dipartimento di Fisica, Universit\`a di Milano, Via
Celoria, 16 - 20133 Milano (Italy)}
\email{angelo.loinger@mi.infn.it}

\vskip0.50cm

\begin{abstract}
The full relativity of the concepts of motion and rest, which is
characteristic of the Einsteinian general relativity (GR), does
\emph{not} allow the generation of \emph{physical} gravitational
waves (GW's). -- The undulatory nature of a metric tensor is
\emph{not} an invariant property, but depends on the coordinate
frame. -- An undulation of a metric tensor is propagated with a
speed that can have any value between zero and infinite.
\end{abstract}

\maketitle


theory.

\normalsize

\vskip1.20cm \noindent \textbf{1}. -- The exact (non-approximate)
formulation of general relativity (GR) does \emph{not} allow the
existence of \emph{physical} gravitational waves (GW's). I have
given several proofs of this fact \cite{1}. Quite simply, we can
observe, \emph{e.g.}, that bodies which interact only
gravitationally describe \emph{geodesic} lines, and therefore --
as it is very easy to see -- they do not generate any GW. If we
add \emph{non}-gravitational forces, the conclusion remains
\emph{the same}, because the new trajectories do \emph{not}
possess kinematical elements (velocity, acceleration, time
derivative of the acceleration, \emph{etc}.) different from those
of the geodesic motions.

\par Another plain proof of the unreality of the GW's runs as
follows. Remember \emph{in primis} that, contrary to what happens
in Maxwell theory, for which the class of the inertial systems has
a physical privilege, in GR \emph{all} coordinate systems are on
the same footing, none of them is physically privileged. Let us
consider, to be determinate, the Schwarzschild solution to
Einstein equations which gives the field generated by a
gravitating spherosymmetrical body $B$ -- as it is seen by an
observer $\Omega$ at rest together with $B$. Now, the two
instances: \emph{i}) body $B$ at rest and observer $\Omega$ in any
whatever motion $W$, and \emph{ii}) $\Omega$ at rest and $B$ {}
$W$-moving, are indistinguishable, because \emph{in GR} we can
only speak of \emph{relative} motions. But $B$ at rest cannot emit
GW's. We can also say: the gravitational potential of $B$ at rest
is characterized by the static $g_{jk}(\textbf{x})$ of
Schwarzschild solution. The gravitational potential of $B$ in
motion \emph{with respect to $\Omega$} is characterized by a given
time-dependent $g_{jk}^{*}(\textbf{x}^{*},t^{*})$. Of course, the
curvature tensor remains \emph{intrinsically} unaltered by the
transition $(\textbf{x},t)\rightarrow (\textbf{x}^{*},t^{*})$.
Assume now that the motion of $B$ happens in a limited spatial
region $L$. Then, it is possible that at a great distance from $L$
the tensor $g_{jk}^{*}$ has a wavy form, which however would
represent a wave due to the starred coordinates
$\textbf{x}^{*},t^{*}$, \emph{i.e.} an illusive undulation.

\par \emph{\textbf{Conclusion}}: \emph{no motion of $B$ generates}
\textbf{\emph{physical}} \emph{GW's}. This argument is more
straightforward than the reasoning developed in a previous
paper\nopagebreak[4] \cite{2}, in which however it is also
considered the case of the relative motion of \emph{two} bodies
that can be \emph{both} reduced to rest, with respect to an
observer $\Omega$, by a convenient reference system (Weyl).

\vskip0.80cm \noindent \textbf{2}. -- A diffuse, wrong belief
affirms: ``A bar turning round its midpoint has a quadrupole
moment, and therefore it generates GW's.'' Two errors: \emph{i})
the quadrupole formula has been derived from the
\textbf{\emph{linear}} version of GR, which is inadequate to treat
the question of the GW's (Weyl 1944, see \cite{3}); \emph{ii}) a
convenient coordinate change reduces the bar to rest; further, in
the \emph{exact} GR there are no kinematical elements of \emph{any
whatever} motion that are responsible for the emission of GW's.
(The ``bar argument'' has many affectionate supporters; thus, in
order to legitimate the use of linearized approximation, it has
been affirmed that for an observer not very distant from the bar
the fact that somewhere far the metric tensor cannot be
represented by small corrections to Minkowskian tensor is
irrelevant. Now, Weyl \cite{3} has demonstrated that the
gravitational field of the linear version exerts \emph{no} force
on matter, \emph{i.e.} is a ``powerless shadow''!).

\par Analogous considerations can be made for the instance of two
masses in relative oscillation, linked together with a spring.

\par The mere fact that the linearized approximation of GR has a
covariant character \emph{only} with respect to Lorentz
transformations ought to be sufficient for not giving it an
unconditioned credit. Moreover, we can also remark that from a
mathematical standpoint the splitting of metric tensor $g_{jk}$ in
a Minkowskian part $\eta_{jk}$ plus another (``small'') part
$h_{jk}$ has in general a dubious legitimacy.

\vskip0.80cm \noindent \textbf{3}. -- The traditional way of
investigating the question of the GW's consists in attempts to
derive from Einstein field equations a solution provided  with an
undulatory character. The conceptual \textbf{\emph{scheme}} can be
sketched as follows \cite{4}.

\par Consider in Minkowski spacetime a hypothetical scalar field
$S(\textbf{r},t)$, which satisfies d'Alembert equation:

\begin{equation} \label{eq:one}
\nabla^{2} S - \frac{1}{c^{2}} \, \frac{\partial^{2}S}{\partial
t^{2}} = -4\pi \varrho \quad;
\end{equation}

a particular solution, as it is known, is given by the retarded
field:

\begin{equation} \label{eq:two}
S_{ret}(\textbf{r},t) = \int \frac{\varrho \,
(\textbf{r}',t')}{|\textbf{r}'-\textbf{r}|} \, \, \textrm{d}V'
\quad,
\end{equation}

where $ct'=ct-|\textbf{r}'-\textbf{r}|$, and the integration
extends over the volume $V'$ in which $\varrho$ is different from
zero. At large distances from $V'$ the field $S_{ret}$, which
satisfies the homogeneous d'Alembert equation in the region where
$\varrho=0$, has the \emph{asymptotic} form:

\begin{equation} \label{eq:three}
S_{ret}^{(a)}(\textbf{r},t) = \frac{1}{r} \, \, \mu
\left(t-\frac{r}{c}, \, \frac{\textbf{r}}{r}\right) \quad,
\end{equation}

the function $\mu$ being given by:

\begin{equation} \label{eq:four}
\mu \left(t-\frac{r}{c}, \, \frac{\textbf{r}}{r} \right) = \int
\rho \left(\textbf{r}', \, t-\frac{r}{c} + \frac{\textbf{r}' \cdot
\textbf{r}/r}{c}\right) \, \textrm{d}V' \quad.
\end{equation}

The general solution of eq.(\ref{eq:one}) can be written:

\begin{equation} \label{eq:five}
S = S_{ret} + S_{in} \quad,
\end{equation}

or

\begin{equation} \label{eq:six}
S = S_{adv} + S_{out} \quad,
\end{equation}

where: $S_{adv}$ is the advanced field, $S_{in}$ and $S_{out}$ are
the ingoing and outgoing fields, solutions of the homogeneous
d'Alembert equation ($\varrho=0$).

\par The above equations have the \textbf{\emph{same}} formal
structure for \textbf{\emph{all}} the inertial frames of
reference. Consequently, they assure us of the
\textbf{\emph{real}} existence of the waves of our field $S$. This
scheme -- as it is well known -- works very well for Maxwell
electromagnetism, but \emph{in GR matters stand otherwise}. First
of all, as I have previously emphasized, the \emph{linearized}
approach must be discarded, owing to its inadequacy to investigate
the GW's \cite{3}. (And the inadequacy of quadrupole formula
remains also for perturbative refinements of third order in $G$
and fifth order in $v/c$). On the other hand, the application of
the above scheme to \emph{exact} GR -- apart from the analytical
difficulties due to the nonlinearity of Einstein equations -- is
doomed to a failure, as we shall see presently.

\par The relevant computations are executed, of course,  in a
given reference frame, usually in a \emph{harmonic} system of
coordinates, and people have succeeded, in particular, in deriving
perturbative expansions of Einstein equations of various kinds,
mainly described with the adjectives ``post-Newtonian'' and
``post-Minkowskian'' \cite{5}. However, as we know, there exist in
GR \emph{no} coordinate frames that are \emph{physically}
privileged -- not even the harmonic ones (contrary to a firm
conviction of Fock \cite{6}). Now, a property is invariant, and
has consequently a physical meaning, \emph{only} if it holds for
\emph{any} system of coordinates. But the wavy character of a
gravitational field is \emph{not} a property independent of the
coordinate system: a field, which is undulatory in a given
reference frame $F$ \emph{loses} this property with the passage to
a suitable frame $F'$ belonging to an infinite class of similar
systems, that are on the same conceptual footing of $F$. This
simple consideration proves the futility of the traditional
attempts to demonstrate the existence of physical GW's. It is also
useful to remark that, dependently on the reference frame, the
speed of an undulation of metric tensor can have \emph{any} value
between zero and infinite, contrary to an old conviction (derived
from the linearized version of GR), which attributes to the GW's
the speed of light $c$ \emph{in vacuo}.

\pagebreak
\begin{center}
\noindent \small \emph{\textbf{APPENDIX A}}
\end{center}
\normalsize \noindent
\par The Einsteinian thesis that in GR there
are no coordinate systems which are physically privileged gave
origin in past times to a lively debate, provoked by the
publication in 1917 of an important article by Kretschmann
\cite{7}. A point on the question was made by Pauli, who showed
convincingly the correctness of Einstein's thesis \cite{8}. Fock
was not persuaded, and always affirmed that in GR too (as in SR)
there is a class of physically privileged frames: the harmonic
ones \cite{9}; but the proof of this assertion is defective.
\emph{Et pour cause}.

\par Back to Kretschmann \cite{7}. His paper was admirably
summarized by Ph. Frank in the following terms \cite{10}:
 ``Einstein versteht unter seinem
allgemeinen Re1ativit\"{a}tsprinzip die Forderung, da\ss {} die
Naturgesetze durch Glei\-chungen ausgedr\"{u}ckt werden sollen,
die gegen\"{u}ber beliebigen Koordinatentransformationen kovariant
sind. Der Verf. zeigt nun, da\ss {} jede beliebigen Gesetzen
gehorchende Naturerscheinung durch allgemeine kovariante
Gleichungen beschrieben werden kann, da\ss {} also das Bestehen
solcher Glei\-chungen keine physikalische Eigenschaft aussagt. Zum
Beispiel kann die gleich\-m\"{a}ssige Ausbreitung des Lichtes im
schwerelosen Raum auch kovariant dargestellt werden. Es ergibt
aber dann eine Darstellung derselben Erscheinungen, die nur eine
engere Gruppe (die Lorentz-Trasformationen) zul\"{a}sst. Diese
Gruppe, die durch keine Darstellung der Erscheinung mehr eingeengt
werden kann, ist f\"{u}r das betreffende System charakteristisch.
Die Invarianz ihr gegen\"{u}ber ist eine physikalische Eigenschaft
des Systems und stellt im Sinne des Verf. das
Relativit\"{a}tspostulat f\"{u}r das betreffende
Erscheinungsgebiet dar. In der Einsteinschen allgemeinen
Relativit\"{a}tstheorie k\"{o}nnen nun durch geeignete Wahl der
Koordinaten die Feldgleichungen auf eine Gestalt gebracht werden,
die nicht mehr gegen\"{u}ber der Gruppe der
Koordinatentransformationen kovariant ist. Der Verf. gibt eine
Reihe von Beispielen solcher Umformungen. Die so umgeformten
Gleichungen gestatten aber \"{u}berhaupt keine Gruppe mehr, und in
diesem Sinne ist die Einsteinsche allgemeine
Relativit\"{a}tstheorie eine ``Absoluttheorie'', w\"{a}hrend die
spezielle Relativitivit\"{a}stheorie auch im Sinne des Verf. dem
Re1ativit\"{a}tspos\-tulat f\"{u}r die Lorentz-Transformationen
gen\"{u}gt.''

\par However, Kretschmann's interpretation of GR, according to
which GR would be in reality an ``Absoluttheorie'', revealed
itself as sterile, and Pauli \cite{8} emphasized that the attempts
by Kretschmann and Mie to ``normalize'' with suitable criteria the
coordinate system resulted fruitful only in special instances.
\emph{In the general case, and in fundamental questions, the
general covariance is indispensable}. Indeed, the general
covariance is an \emph{essential} and \emph{characterizing}
property of GR, while for other physical theories (Maxwell
electrodynamics, special relativity, \emph{etc}.) the formulation
in terms of general coordinates is only an \emph{additional}
possibility, which leaves unchanged their physical
characteristics.

\par In a sense, the astrophysical community considers GR as an
``Absoluttheo\-rie'' in regard to the class of the cosmological
models, with their ``cosmic times''. The instance of Friedmann
models -- which however are isomorphic to corresponding Newtonian
models -- is very significative.

\par In reality, we should bear in mind that \emph{not} all the
properties attributed to a given cosmological model have an
\emph{invariant} character, and are therefore \emph{real}
properties. In particular, the age of the universe \emph{depends}
on the coordinate system that was chosen for the considered model.
Further, the durations of the temporal stages in which the above
age is usually divided in the instance of a Friedmann model, have
only a \emph{conventional} value from the general-relativistic
standpoint. Only if we consider the isomorphic \emph{Newtonian}
model, all time periods have an absolute meaning.

\vskip0.80cm
\begin{center}
\noindent \small \emph{\textbf{APPENDIX B}}
\end{center}

\normalsize \noindent
\par In April, in May, and in June 2007 three teams of
astrophysicists published the following three papers:

\begin{itemize}
\item[-] ``Maximum Entropy for Gravitational Waves Data Analysis:
Inferring the Physical Parameters of Core-Collapse Supernovae''
\cite{11}; \item[-] ``Rates and Characteristics of Intermediate
Mass-Ratio Inspirals Detectable by Advanced LIGO'' \cite{12};
\item[-] ``Host Galaxies Catalog Used in LIGO Searches for Compact
Binary Coalescence Events'' \cite{13}.
\end{itemize}

\par The authors describe various astrophysical phenomena that,
in their opinion, should generate GW's, and could be detectable by
LIGO interferometers. It is easy to foresee that no GW will be
registered by the apparatuses.

\end{document}